\documentclass{aipproc}

\newcommand{\lmin}{L_{\rm min}}
\newcommand{\lmax}{L_{\rm max}}

\def\mathnew{\mathsurround=0pt}
\def\simov#1#2{\lower .5pt\vbox{\baselineskip0pt \lineskip-.5pt
        \ialign{$\mathnew#1\hfil##\hfil$\crcr#2\crcr\sim\crcr}}}

\layoutstyle{8x11double}

\title{Determining the GRB (Redshift, Luminosity)-Distribution Using
Burst Variability}

\author{Timothy Donaghy}
{
address={Department of Astronomy \& Astrophysics, University of Chicago,
5640 South Ellis Avenue, Chicago, IL 60637},
email={lamb@oddjob.uchicago.edu},
}

\author{Donald Q. Lamb}
{
address={Department of Astronomy \& Astrophysics, University of Chicago,
5640 South Ellis Avenue, Chicago, IL 60637},
email={lamb@oddjob.uchicago.edu},
}

\author{Daniel E. Reichart}
{
address={Department of Astronomy, California Institute of Technology,
1201 East California Boulevard, MS 105-24, Pasadena, CA 91125},
email={der@astro.caltech.edu},
}

\author{Carlo Graziani}
{
address={Department of Astronomy \& Astrophysics, University of Chicago,
5640 South Ellis Avenue, Chicago, IL 60637},
email={lamb@oddjob.uchicago.edu},
}

\begin{abstract}
We use the possible Cepheid-like luminosity  estimator for the
long-duration gamma-ray bursts (GRBs) developed by Reichart et al.
(2000) to estimate the intrinsic luminosity, and thus the redshift, of
907 long-duration GRBs from the BATSE 4B catalog.   We describe a
method based on Bayesian inference which allows us to infer the
intrinsic GRB burst rate as a function of redshift for bursts with
estimated intrinsic luminosities and redshifts. We apply this method to
the above sample of long-duration GRBs, and present some preliminary
results.
\end{abstract}

\begin{document}

\maketitle

\section{Introduction}

There is increasing evidence that gamma-ray bursts (GRBs) are due to
the collapse of massive stars (see, e.g., [1] for a discussion of this
evidence).  If GRBs are indeed related to the collapse of massive
stars, one expects the GRB rate to be roughly proportional to the
star-formation rate (SFR).  However, the observed redshift distribution
of GRBs differs noticeably from that of the SFR:  the observed GRB
redshift distribution peaks at $z \approx 1$ and few bursts are
observed beyond $z\sim 1.5$, while the SFR peaks at $z \approx 2$ and
10-40\% of stars are thought to form beyond $z=5$ (see, e.g., [2,3]).

\begin{figure}[ht]
\includegraphics[scale=0.35]{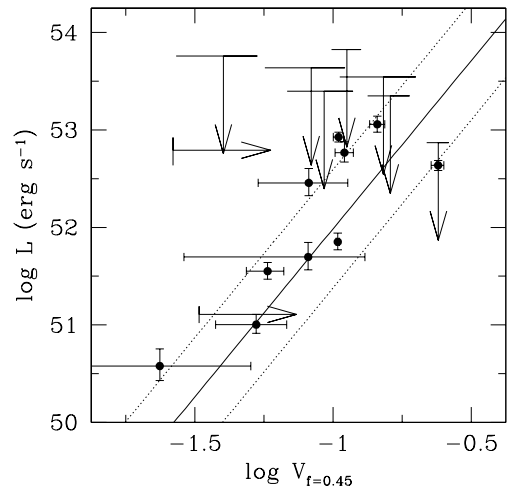}
\caption{The variabilities $V$ and isotropic-equivalent peak photon
energy luminosities $L$ of the 19 bursts for which some redshift
information exists.  The solid and dotted lines mare the center and
1-$\sigma$ widths of the best-fit model of these bursts in the (log
$L$, log $V$)-plane.}
\label{F:tdonaghy:1}
\end{figure}

However, observational selection effects play a key role in determining
the observed redshift distribution of GRBs.  Among these selection
effects are the efficiencies with which burst redshifts can be
determined by spectroscopic observations of the burst afterglow and/or
the host galaxy of the burst.  In addition, both of these methods of
determining redshift require the identification of an optical afterglow
of the GRB, and the detectability of the optical afterglow may be a
function of redshift.

\begin{figure}[t]
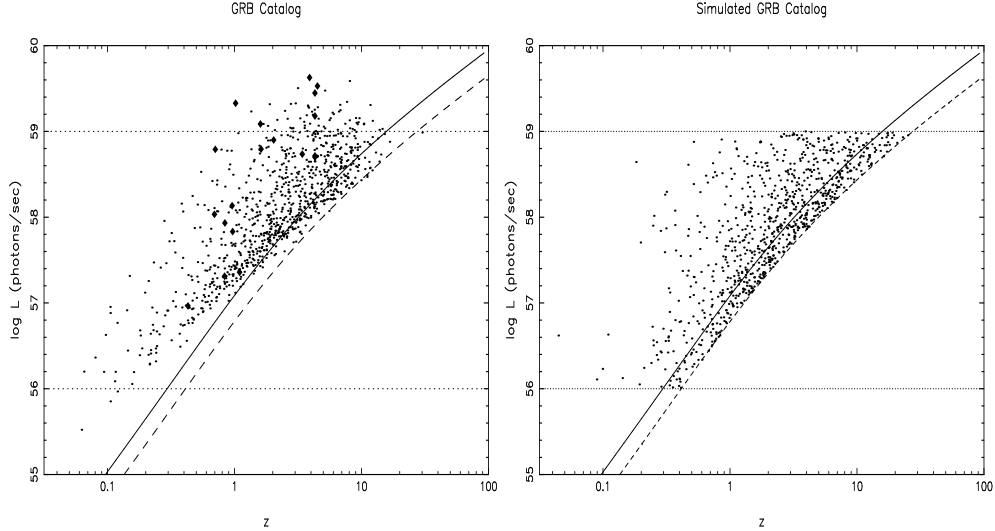
 
%\begin{center}
\rotatebox{270}{\resizebox{7cm}{6.5cm}{\includegraphics{lumest.ps}}}
\rotatebox{270}{\resizebox{7cm}{6.5cm}{\includegraphics{best_fit_noevol_p90.zL.ps}}}
%\end{center} 
\caption{Left panel: distribution of the 907 long-duration BATSE bursts
in the (log $z$, log $L$)-plane, as determined by our luminosity
estimator.  The solid and dashed curves represent the 90\% and 10\%
detection thresholds of BATSE.  The horizontal dotted lines show the
values of $\lmin$ and $\lmax$ adopted in this study.  The diamonds show
the locations of the 19 GRBs with known redshifts used to calibrate the
luminosity estimator.
Right panel: a Monte Carlo simulation of the expected distribution of
$\approx$ 900 bursts with peak photon fluxes $P$ above the BATSE 90\%
detection threshold, assuming the maximum likelihood best-fit
parameters for the model (see text).}
\label{F:tdonaghy:2} 
\end{figure}

\begin{figure}[t]
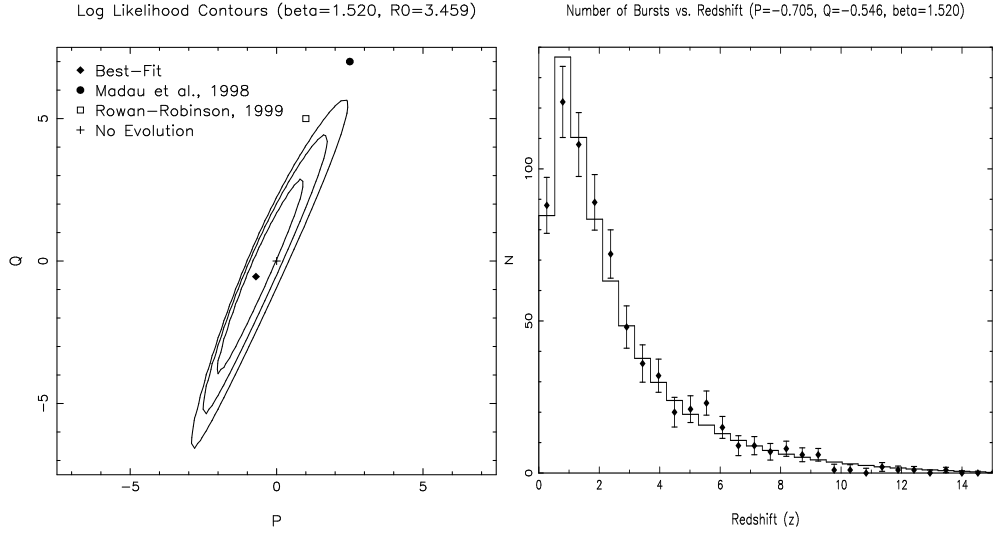
 
\rotatebox{270}{\resizebox{7cm}{6.5cm}{\includegraphics{cont1_p90.ps}}}
\rotatebox{270}{\resizebox{7cm}{6.5cm}{\includegraphics{comp_noevol_p90.ps}}}
\caption{Left panel: the best-fit point in the ($P, Q$)-plane at  the
maximum likelihood best-fit values of the normalization $R_0 = 3.46$
of the GRB rate and the power-law index $\beta =1.52$ of the GRB
luminosity distribution.  The contours shown correspond to $\Delta \log
{\cal L} = 150$.  Right panel:  a comparison of the expected and observed
number of bursts with $P$ above the BATSE 90\% detection threshold for
the maximum likelihood best-fit model.}
\label{F:tdonaghy:3}
\end{figure}

\begin{figure}[t]
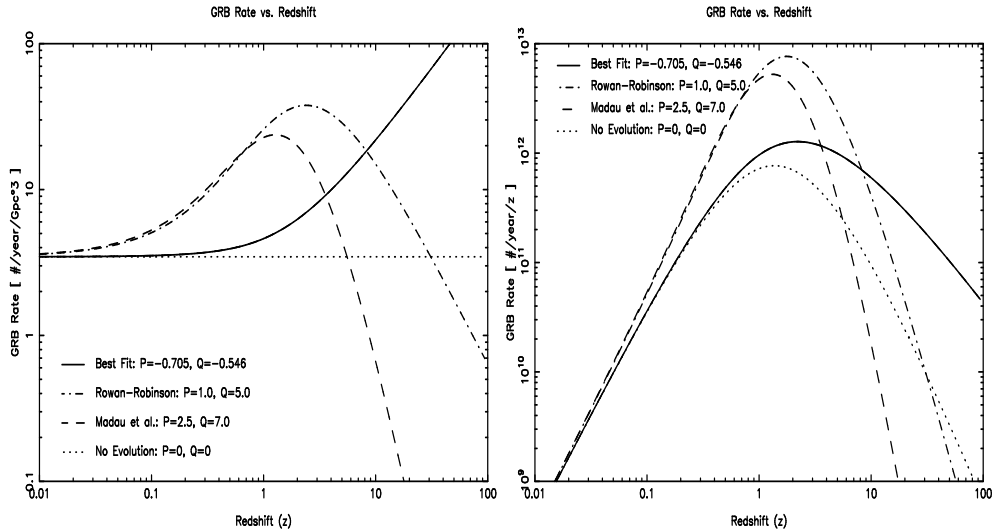
 
\rotatebox{270}{\resizebox{7cm}{6.5cm}{\includegraphics{rates_vol_p90.ps}}}
\rotatebox{270}{\resizebox{7cm}{6.5cm}{\includegraphics{rates_z_p90.ps}}}
\caption{Left panel: the best-fit GRB rate per unit comoving volume
(solid line).  Right panel: the GRB rate per unit redshift $z$. 
Estimates of the star formation rate as a function of redshift $z$ made
by Madau et al. (1998) and Rowan-Robinson (1999) are shown in both
panels for comparison, as is the no evolution with redshift $z$ model
($P=Q=0$).  We emphasize that we have not taken into account the
statistical and systematic errors in the redshifts $z$ and intrinsic
peak photon luminosities $L$ derived from the variability measure $V$,
and therefore cannot quote meaningful confidence regions for our
best-fit  parameters.
}
\label{F:tdonaghy:4}
\end{figure}

In this paper, we take a different approach.  We use the possible 
Cepheid-like luminosity estimator for the long-duration GRBs developed
by Reichart et al. [4] to estimate the intrinsic luminosity, and thus
the redshift, of 907 long-duration GRBs from the BATSE 4B catalog [5]. 
This approach is free of the important and difficult-to-quantify
observational selection effects that affect the redshift distribution
of the GRBs with known redshifts.  We assume a very general model for
the GRB rate and a truncated power-law model for the intrinsic GRB
isotropic-equivalent photon luminosity distribution (in this work we
assume that the amplitude and the power-law index of the
isotropic-equivalent photon luminosity distribution does not evolve; we
will relax this assumption in future work).  Adopting a Bayesian
approach, we calculate the likelihood of the data given the model, and
convert it to a posterior distribution on the model parameters.

\section{Luminosity Estimator}

A possible Cepheid-like luminosity estimator for gamma-ray bursts has
been suggested by Ramirez-Ruiz and Fenimore [6] and developed further 
by Reichart et al. [4,7,8].  These authors have shown that there exists
a correlation between a measure $V$ of the variability of the burst
time history and the intrinsic isotropic-equivalent peak photon energy
luminosity $L$ of the burst for the 19 GRBs for which some redshift
information (either a redshift measurement or a redshift limit)
exists.  We show this correlation in Figure \ref{F:tdonaghy:1}.

We apply this luminosity estimator to 907 bursts from the BATSE 4B 
catalog [5] that have durations $T_{90}> 10$ sec.  We show the
resulting redshift-luminosity distribution for these bursts in Figure
\ref{F:tdonaghy:2} (see also [7] and [8]).

\section{Model}

In this work, we assume that the rate of GRBs per unit redshift and
luminosity is given by a separable function of redshift $z$ and
intrinsic isotropic-equivalent photon number luminosity $L$.  That is,
we assume that the intrinsic photon number luminosity distribution of
GRBs does not evolve with redshift; we will relax this assumption in
future work.  Then the rate of GRBs per unit redshift and luminosity 
can be written as [9]
\begin{equation}\label{E:tdonaghy:model}
\frac{dN}{dz\,dL_N}=\rho(z)f(L) \; ,
\end{equation}
\vskip -5pt
\noindent
where
\begin{equation}
\rho(z)=R_{\rm GRB}(z;P,Q) \times (1+z)^{-1}\times 4\pi r(z)^2 (dr/dz)
\end{equation}
is the rate of GRBs that occur at redshift $z$, and $r(z)$ is the
comoving distance to the source.

We adopt a phenomenological model for the rate of GRBs that occur at 
redshift $z$ per unit comoving volume of the form proposed by
Rowan-Robinson for the star formation rate [10].  In this model, the
GRB rate is given by
\begin{equation}\label{E:tdonaghy:rateGRB}
R_{\rm GRB}(z;P,Q) = R_0 \left( \frac{t(z)}{t(0)} \right)^{P} 
	\exp\left[ -Q\left( 1-\frac{t(z)}{t(0)} \right)\right].
\end{equation}
where $P$, $Q$ and $R_0$ are model parameters, and $t(z)$ is the time
since the Big Bang corresponding to the redshift $z$.

We take the intrinsic photon luminosity distribution of GRBs to be a
truncated power-law,  
%\begin{equation}\label{E:tdonaghy:lumGRB}
\vskip -12pt
\begin{eqnarray}
f(L) &=& {{1-\beta} \over {L^{1-\beta}_{\rm max}-L^{1-\beta}_{\rm min}}} 
L^{-\beta}\nonumber \\
&& \times \Theta(L-\lmin) \Theta(\lmax-L).
\end{eqnarray}
%\end{equation}
Thus the model has six parameters:  $P$, $Q$, $R_0$, $\beta$, $\lmin$,
and $\lmax$.

In this work we assume a flat universe with $\Omega_{\rm M}=0.3$ and
$\Omega_{\Lambda}=0.7$.  Furthermore, we fix $\lmin$, and $\lmax$ to be
constants (see Figure \ref{F:tdonaghy:2}), reducing the number of free
parameters in our model to four.

\section{Likelihood Function}

The likelihood function for GRBs that are Poisson in time is given by
[9]
\begin{equation}
{\cal L}=\exp\left\{-\int dz\,dP\,\mu(z,P)\epsilon(z,P)\right\}
\prod_{i=1}^N \mu(z_i,P_i) \; ,
\end{equation}
\noindent
where the event likelihood,
\vskip -12pt
\begin{eqnarray}
\mu(z_i,P_i)&=&\int_0^\infty dL\,\rho(z_i)f(L)
\delta\left(P_i-{{L} \over {4\pi r(z_i)^2(1+z_i)^\alpha}}\right)\nonumber \\
&=&\rho(z_i)\times f\left(4\pi r(z_i)^2(1+z_i)^\alpha P_i\right)\nonumber \\
&&\times 4\pi r(z_i)^2(1+z_i)^\alpha \; ,
\end{eqnarray}
is the expected number of events observed within $dz\,dP$ of redshift
$z_i$ and peak photon number flux $P_i$.  The quantity $\alpha$ is the
burst spectral index,  which we take to be equal to $1.0$, a value that
is typical of GRBs [11].  By an application of Bayes' Theorem, we
regard ${\cal L}$ as an (unnormalized) probability distribution on the
model parameters.  We estimate the best-fit parameters in our model by
maximizing this likelihood function over the parameter space.

We take the BATSE observing efficiency $\epsilon(z,P)$, which appears
in the exponential factor in the likelihood function, to be
$\theta(P-P_{th})$, where $P_{th}=0.4$ photons cm$^{-2}$ sec$^{-1}$ is
the $90$\% BATSE detection threshold.  The product of the event
likelihoods $\mu(z_i,P_i)$ runs over the 907 BATSE bursts considered in
this study.

\section{Results}

Each ($z,L$)-point in the left panel of Figure \ref{F:tdonaghy:2}
corresponds to the maximum value of a probability distribution for that
particular burst. Thus each point has statistical and systematic errors
that are associated with it and that are not displayed in the  figure. 
Since the estimate of the redshift $z$ is calculated from the estimate
of the peak photon luminosity $L$ (and the peak observed photon number
flux $P$), the errors in $z$ and $L$ are completely correlated; that
is, the uncertainty lies along curves of constant $P$ (i.e., they are
diagonals) in the ($z, L$)-plane.  Furthermore, these uncertainties are
not symmetric; they are skewed toward low $L$ at low $z$ and toward
high $L$ at high $z$ [7,8].

In this preliminary work, we neglect these statistical and systematic
errors, and use only the best-fit values in our analysis.  Calculating
the likelihood of this data, given the model, and converting it to a
posterior distribution on the model parameters, we find the maximum
likelihood parameters for this data set to be: $P=-0.70$, $Q=-0.55$,
$R_0 = 3.46$ and $\beta = 1.52$.  The right panel of Figure
\ref{F:tdonaghy:2} shows a Monte Carlo simulation of the distribution
of $\approx$ 900 bursts whose peak photon fluxes $P$ are above the
BATSE 90\% detection threshold for the maximum likelihood best-fit
parameters.

The left panel of Figure \ref{F:tdonaghy:3} shows the best-fit point
and contours of $\Delta \log {\cal L} = 150$ in the ($P, Q$)-plane at
the maximum likelihood best-fit values of the normalization $R_0 =
3.46$ of the GRB rate and the power-law index $\beta =1.52$ of the GRB
luminosity distribution.  The left panel of Figure \ref{F:tdonaghy:3}
thus represents a two-dimensional slice through the four-dimensional
parameter space of the model.  The contours shown in Figure
\ref{F:tdonaghy:3} do not correspond to credible regions because we
have not taken into account the statistical and systematic errors in
the redshifts $z$ and intrinsic peak photon luminosities $L$ derived
from the variability measure $V$.  However, taking the preliminary
results at face value, the best-fit model appears to be consistent with
no evolution but to be inconsistent with the Rowan-Robinson [10] and
the Madau et al. [12] expressions for the star formation rate as a
function of $z$.

The right panel of Figure \ref{F:tdonaghy:3} shows a comparison of the
observed and expected numbers of bursts with $P$ above the BATSE 90\%
detection threshold for the maximum likelihood best-fit model.  These
distributions thus correspond to projections onto the $z$-axis of the 
burst distributions in the left and right panels of Figure 
\ref{F:tdonaghy:2}, respectively.

The left panel of Figure \ref{F:tdonaghy:4} shows the best-fit GRB rate
per unit comoving volume, while the right panel shows the GRB rate per
unit $z$.  Estimates of the star formation rate as a function of $z$
made by Madau et al. (1998) and Rowan-Robinson (1999) are shown for
comparison, as is the no evolution model ($P=Q=0$).

Applying the intrinsic peak luminosity estimator of Ramirez-Ruiz and
Fenimore [5] to the brightest 220 long-duration bursts in the BATSE 4B
catalog [11], Lloyd-Ronning et al. [13] find that the GRB rate
increases with increasing redshift ($R_{\rm GRB} \propto (1+z)^3$ until
$z \sim 2$ and $\propto (1+z)^1$ from $z \sim 2$ until $z \sim 10$),
and that the mean luminosity of the GRBs also increases with increasing
redshift ($<L> \propto (1+z)^{1.4 \pm 0.2}$).  The results found in the
present preliminary work differ somewhat from these conclusions.  But
we again emphasize that (1) the model that we have used in the present
preliminary study does not allow for the evolution of $L$ with $z$, and
(2) we have not taken into account the statistical and systematic
errors in the redshifts $z$ and intrinsic peak photon luminosities $L$
derived from the variability measure $V$, and therefore cannot quote
meaningful confidence regions for our best-fit parameters.  In future
work, we will relax the assumption of no evolution and will take into
account the statistical and systematic errors.

{\voffset -12pt}

\end{document}